\documentclass[10pt,twocolumn]{article}

\usepackage[a4paper,margin=1.75cm]{geometry}
\usepackage{amsmath,amssymb,booktabs,graphicx,microtype}
\usepackage{balance}
\usepackage{xcolor} 
\usepackage[font=small,labelfont=bf]{caption}
\usepackage{float}
\usepackage[hidelinks]{hyperref}
\usepackage[numbers,sort&compress]{natbib}
\providecommand{\doi}[1]{\href{https://doi.org/#1}{doi:\nolinkurl{#1}}}


\newcommand{\zframe}{\mathbf Z}
\newcommand{\xframe}{\mathbf x}
\newcommand{\kbt}{k_{\mathrm B}T}

\title{MD-LLM-2: A Transferable Language Model of Molecular Dynamics with Physical Conditioning and Explicit Path Probabilities}

\author{Mhd Hussein Murtada$^{1}$, Z. Faidon Brotzakis$^{2,3}$, and Michele Vendruscolo$^{1,*}$\\
{\small $^{1}$Yusuf Hamied Department of Chemistry, University of Cambridge, Cambridge, UK}\\
{\small $^{2}$Department of Biotechnology, Agricultural University of Athens, 75 Iera Odos Str., 11855 Athens, Greece}\\
{\small $^{3}$Archimedes Unit, Athena Research Center, 15125 Marousi, Greece}
}
\date{}

\begin{document}
\maketitle

\begingroup
\renewcommand{\thefootnote}{*}
\footnotetext{Corresponding author: mv245@cam.ac.uk}
\endgroup

\begin{abstract}
Molecular dynamics (MD) simulations model the motion of proteins between conformational states, but crossing free energy barriers and sampling multiple thermodynamic conditions remain computationally costly. We recently introduced MD-LLM-1, which showed that language models can generate conformational trajectories at lower cost, but required a separate model for each protein and did not explicitly model thermodynamics or kinetics. Here we introduce MD-LLM-2, a transferable language model that generates trajectories conditioned on sequence, temperature and molecular history. Because each step defines a normalized distribution over structural tokens, the model assigns an explicit likelihood to every generated or simulated trajectory. Trained across mdCATH domains, MD-LLM-2 reproduces the median radius of gyration of MD within 1~\AA{} in 10 of 12 unseen domains, predicts reduced Trp-cage compaction at higher temperature, and generates L99A T4 lysozyme trajectories that reach published transition-region backbone structures. In alanine dipeptide simulations, conditioning on a torsional scaling parameter $\lambda$ predicts basin populations at unseen couplings with a mean absolute error of 0.09 versus 0.35 for MBAR, and first-passage probabilities with an error of 0.08 versus 0.16 for a reversible Markov state model. Frame-isoentropic sampling further limits the accumulation of errors during autoregressive trajectory generation. Together, these results extend MD-LLM from protein-specific trajectory generation to transferable, physically conditioned modelling with explicit path likelihoods, and show that the framework can predict equilibrium populations for force-field parameters not seen during training, with an initial indication that first-passage kinetics can also be captured.
\end{abstract}

\section{Introduction}

The conformational populations of proteins and the rates of interconversion between them depend on the physical conditions of the system and on the underlying molecular interactions. Resolving these changes with molecular dynamics (MD) is computationally demanding, because atomistic simulations require femtosecond integration timesteps, whereas conformational transitions may occur on microsecond-to-millisecond timescales, if not longer \cite{shaw2010atomic,hollingsworth2018molecular,shaw2021anton3}. Enhanced sampling methods such as metadynamics and Hamiltonian replica exchange accelerate the sampling of rare states \cite{laio2002metadynamics,fukunishi2002hrex,chen2022enhancing,ray2023kinetics}, but they remain costly and system-specific, relying on predefined collective variables or on a ladder of simulated thermodynamic states.

Data-driven models provide an alternative strategy by learning statistical descriptions of molecular ensembles and dynamics from existing data. Markov state models and VAMPnets infer conformational kinetics from molecular trajectories \cite{prinz2011msm,mardt2018vampnets}, whereas BioEmu generates conformational ensembles approximating equilibrium distributions across proteins, without explicit time evolution \cite{lewis2025bioemu}. TITO learns finite-lag transition densities for small molecules and peptides that generalise across molecular systems and lag times, and has been shown to recover equilibrium probabilities and slow relaxation dynamics at long lags \cite{diez2026tito}. PLaTITO extends this framework to coarse-grained proteins, training on mdCATH trajectories across multiple temperatures and using protein language model embeddings to improve generalisation; it reports state-of-the-art out-of-distribution equilibrium sampling and captures folding kinetics \cite{antoniadis2026platito}. Both represent finite-lag transition laws with continuous generative models, for which transition likelihoods, although defined, typically require integrating a probability-flow ODE at each step. 

MD-LLM takes a different approach to learning molecular dynamics, based on language modelling \cite{murtada2024language,murtada2025mdllm1}. At a conceptual level, structural tokens play the role of words, molecular conformations the role of sentences, and trajectories the role of documents: each conformation is encoded as a sequence of discrete structural tokens, and a large language model (LLM) learns how one conformation follows another from the preceding molecular history. Learning molecular dynamics is thus recast as a generative sequence-modelling problem, in which the statistical relationships between successive conformations are learned from simulation trajectories. Because every token is drawn from a normalised categorical distribution, the probability of each conformational transition, and hence of a complete tokenised trajectory, follows directly from the product of these probabilities, without the numerical integration required to evaluate densities in continuous normalising-flow models.

MD-LLM-1 introduced this approach for individual proteins \cite{murtada2025mdllm1}, generating trajectories encoded with FoldToken4 \cite{gao2024foldtoken4} that explored alternative conformational regions in T4 lysozyme (T4L) and Mad2
\cite{murtada2025mdllm1}. More recently, ProTDyn developed a closely related autoregressive language-model formulation based on discrete protein structural tokens (ESM3), applying it to transferable equilibrium and multi-timescale trajectory generation and highlighting the availability of exact model likelihoods \cite{liu2025protdyn}. MD-LLM-2 differs in explicitly conditioning trajectory generation on physical variables, including temperature and force-field parameters, in adapting a general-purpose LLM, and in using the model's predictive uncertainty to control autoregressive sampling.

Building on MD-LLM-1, MD-LLM-2 extends the original approach from protein-specific models to a transferable model trained across protein domains, retains frame-by-frame autoregressive generation with ESM3 structural tokens \cite{hayes2025esm3}, and introduces conditioning on physical variables such as temperature and Hamiltonian coupling. We formulate the coordinate-space target as a history-conditioned finite-lag transition distribution
\begin{equation}
p_{\Delta t}(\xframe_{t+\Delta t}\mid H_t^x,S,T,\lambda),
\label{eq:transfer}
\end{equation}
where $H_t^x$ is a history of recent configurations, $S$ is the sequence, $T$ is the physical temperature and $\lambda$ is an optional Hamiltonian coupling. The lag $\Delta t$ specifies the horizon over which a new configuration is predicted. In a reduced representation, the Mori--Zwanzig formalism shows that the projected dynamics generally retain memory of earlier configurations \cite{zwanzig1961memory,mori1965transport}. For models conditioned only on the current reduced state, a Markovian description is most appropriate at lags for which unresolved memory is sufficiently weak. TITO and PLaTITO \cite{diez2026tito,antoniadis2026platito} adopt such finite-lag Markovian transition formulations. MD-LLM-2 can instead condition explicitly on a history of recent frames, allowing some of this memory to be represented rather than absorbed solely into the choice of lag. The same autoregressive formulation can therefore be applied at the 0.2-ps lag used for alanine dipeptide and at the 1-ns stored-frame lag of mdCATH \cite{mirarchi2024mdcath}.

A tokeniser maps each configuration $\xframe_t$ to $N$ structural tokens $\zframe_t=(z_{t,1},\ldots,z_{t,N})$. We index token frames by step, so that $\zframe_{t+1}$ encodes the configuration one lag later, and collect the conditioning variables $S$, $T$ and, where used, $\lambda$ in $c$. Given the tokenised history $H_t$, the model generates the next frame one token at a time:
\begin{equation}
q^{\rm tr}_\theta(\zframe_{t+1}\mid H_t,c) =\prod_{r=1}^{N}q^{\rm tr}_\theta (z_{t+1,r}\mid H_t,c,z_{t+1,<r}),
\label{eq:ar}
\end{equation}
where $z_{t+1,<r}$ denotes the tokens of the next frame generated before position $r$. Each token is drawn from a normalized categorical distribution over the structural codes, so for a fixed starting frame and sampling policy, the model defines a normalised distribution over token paths. Three properties follow. The likelihood of any trajectory, generated or from MD, can be evaluated directly. When the token space is small, the distribution can be enumerated and state populations summed exactly, without sampling. Finally, the predictive distribution at each step exposes the model's uncertainty, which can be used to guide sampling.

We introduce frame-isoentropic sampling, which uses this uncertainty to limit rollout drift arising from exposure bias, the accumulation of errors in autoregressive generation once generated outputs replace training data in the history \cite{bengio2015scheduled,ranzato2016sequence}. In T4L rollouts, the mean categorical entropy of the model's predictions rises with structural drift (median within-rollout Spearman $\rho=0.855$, Supplementary Results). After each generated frame $s$, a controller selects one sampling temperature $\tau_{s+1}$ for all tokens of the next frame, so that their mean predictive entropy $h_s(\tau_{s+1})$ returns to a reference value $h_\star$ set from probes of the first transition. This feedback is conceptually related to the velocity-rescaling thermostat of Bussi, Donadio and Parrinello
\cite{bussi2007velocity}; here it acts on token probabilities while the
physical $T$ and $\lambda$ remain fixed. Feedback-based sampling has also been used to control the perplexity of generated text \cite{basu2021mirostat}.

Overall, we show here that MD-LLM-2 is a transferable, physically conditioned language model of molecular trajectories. Trained across mdCATH domains, the model generates trajectories for proteins absent from training, predicts reduced Trp-cage compaction at higher temperature, and reaches backbone conformations close to published structures of the transition region between the ground and excited states of L99A T4L \cite{vallurupalli2016atomistic,bouvignies2011excited}. Low-populated states of this kind are central to protein function and misfolding but are difficult to access by simulation, and have been characterised at atomic resolution by relaxation dispersion NMR \cite{korzhnev2004fyn,neudecker2012intermediate,bouvignies2011excited}. Throughout, conditioning on molecular history keeps generated paths close to the native scaffold. In alanine dipeptide simulations, Hamiltonian conditioning enables prediction of equilibrium populations at unseen couplings with lower errors than MBAR and, in an initial test of kinetics, first-passage probabilities with lower errors than a $\lambda$-interpolated reversible MSM constructed from the same training simulations. Frame-isoentropic sampling further limits autoregressive drift, reducing the loss of native contacts in the no-history protein model and the largest deviations from MD in unseen domains and T4L. These results show that discrete autoregressive language models can transfer across proteins and respond to physical conditioning while retaining explicit probabilities for molecular trajectories.

\section{Results}

\subsection{Hamiltonian conditioning in alanine dipeptide}

To test Hamiltonian conditioning, we simulated alanine dipeptide in implicit solvent with its torsional energy terms scaled by a coupling $\lambda$ (Eq.~\ref{eq:hamiltonian}), so that $\lambda=1$ recovers the original force field and $\lambda=0$ removes these terms. Training data came from Hamiltonian replica-exchange (HREX) simulations at $\lambda=0.8$--$1.3$ and from fixed-$\lambda$ MD at $\lambda=1.0$, 1.3 and 2.0. Independent references at couplings excluded from training came from a separate HREX ladder spanning $\lambda=0$--$2$ for equilibrium populations and from fixed-$\lambda$ MD for first-passage kinetics. We tested the MD-LLM-2 formulation in three regimes: equilibrium populations below the training range ($\lambda=0.2$--$0.5$), extrapolation to $\lambda=0$, and finite-time kinetics at held-out couplings inside ($\lambda=1.15$) and below ($\lambda=0.8$) the training range.

\begin{figure*}[t]
\centering
\includegraphics[width=0.95\linewidth]{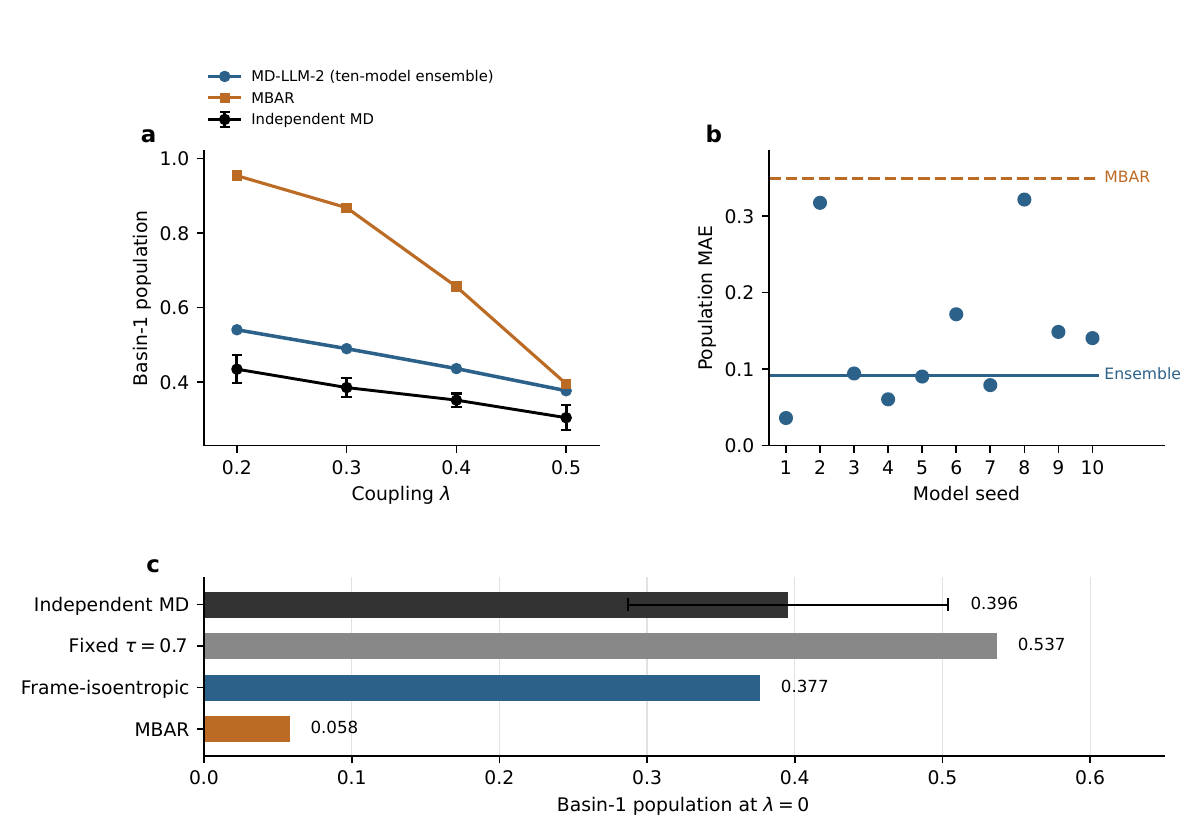}
\caption{\textbf{Hamiltonian-conditioned populations of alanine dipeptide.}
(a) Basin-1 populations at unseen $\lambda=0.2$--$0.5$ from the ten-model MD-LLM-2 ensemble, MBAR reweighting of the training simulations and independent MD. Error bars show 95\% intervals across the eight reference HREX chains. (b) Population mean absolute error of each independently trained model; horizontal lines mark the ensemble and MBAR. Model predictions were fixed before the reference simulations were run. (c) Basin-1 population at $\lambda=0$ over 200 generated frames from an MD-LLM-2 model trained at $\lambda=0.8$--$1.3$, sampled at a fixed $\tau=0.7$ or with frame-isoentropic sampling, compared with MBAR and independent MD. The independent-MD error bar is a 95\% interval across four fixed-$\lambda$ chains.}
\label{fig:alasweep}
\end{figure*}

At $\lambda=0.2$--$0.5$, MD-LLM-2 follows the trend of the reference populations, with a small systematic overestimate, and is closer to them than MBAR. Here MD-LLM-2 predicts single frames conditioned on $\lambda$, and each frame consists of two tokens drawn from 128-code vocabularies, so we enumerated all $128^2$ code pairs, decoded them and summed the probabilities of pairs in basin 1, the $\alpha_R$-like region centred at $\psi\approx-0.92$~rad (Eq.~\ref{eq:population}), at a sampling temperature $\tau=0.7$ selected on training-state validation. Across the four unseen couplings, an ensemble of ten independently trained models reached a mean absolute error of 0.0918, compared with 0.3495 for MBAR reweighting of the same training simulations (Fig.~\ref{fig:alasweep}a,b). Every individual model improved on MBAR, with errors from 0.0358 to 0.3217. MBAR is less effective at these couplings because the training states overlap poorly with the low-$\lambda$ targets. A calibration using two additional MD populations is reported in Supplementary Results.

At $\lambda=0$, frame-isoentropic sampling brings MD-LLM-2 close to the reference population. Using a model trained on the same couplings, the basin-1 population over 200 generated frames was 0.377, compared with 0.396 in independent MD, corresponding to free-energy differences of 0.504 and 0.424~$\kbt$ (Fig.~\ref{fig:alasweep}c). The same model sampled at the fixed $\tau=0.7$ gave 0.537, and MBAR gave 0.058. The controller sets its entropy target from the model's own predictions at $\lambda=0$ and uses no reference data at this coupling.

\begin{figure*}[t]
\centering
\includegraphics[width=0.91\linewidth]{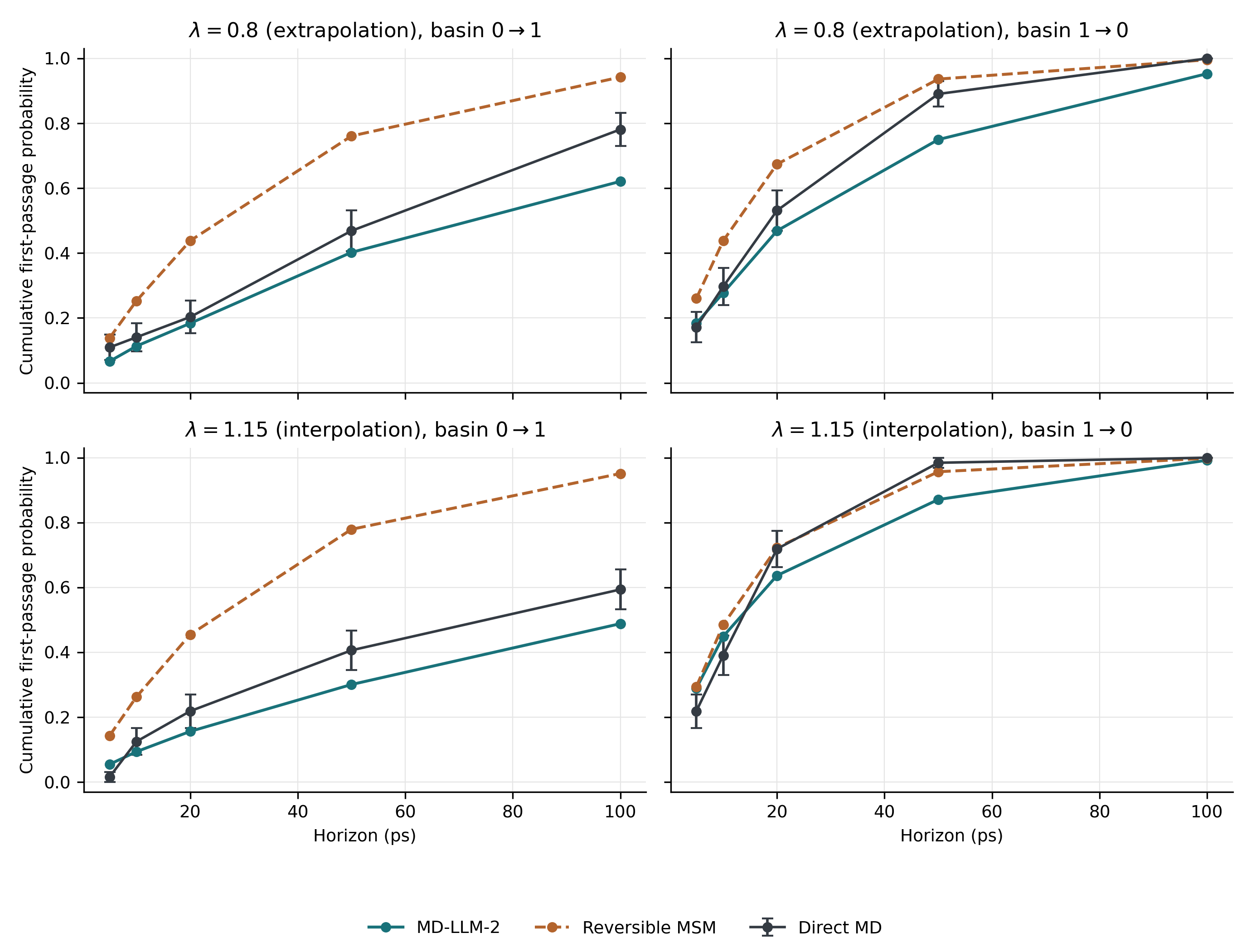}
\caption{\textbf{Finite-time kinetics at held-out Hamiltonian couplings.}
Cumulative first-passage probabilities between basin 0 ($\psi\approx2.75$~rad) and basin 1 ($\psi\approx-0.92$~rad) at $\lambda=0.8$ (below the training range) and $\lambda=1.15$ (within it), from direct MD, MD-LLM-2 and a reversible MSM built from the same training trajectories. Error bars on direct MD show one binomial standard error over 64 paths. The reported mean absolute error uses the 20, 50 and 100-ps horizons. At both couplings, MD-LLM-2 is more accurate for transitions from basin 0 to basin 1, and the MSM for the reverse direction.}
\label{fig:alakinetics}
\end{figure*}

At held-out couplings, MD-LLM-2 also reproduces finite-time kinetics. Generating trajectories autoregressively at a 0.2-ps lag, conditioned on the preceding frames, the model produced 100-ps paths at $\lambda=1.15$ and 0.8, and we compared their cumulative first-passage probabilities between the two $\psi$ basins with direct MD at 20, 50 and 100~ps. The mean absolute error across both couplings, both directions and all three horizons was 0.0811, compared with 0.1568 for a reversible Markov state model (MSM)
\cite{prinz2011msm} built from the same training trajectories (95\%
path-bootstrap interval for the difference, 0.005 to 0.135; Fig.~\ref{fig:alakinetics}). The largest difference between the methods arose for transitions into basin 1, which the MSM predicted to occur too quickly (Supplementary Table S1).

\subsection{Transfer across protein domains and physical temperatures}

We trained MD-LLM-2 on mdCATH trajectories of protein domains up to 200 residues, simulated at five temperatures between 320 and 450~K
\cite{mirarchi2024mdcath}, with trajectories assigned at random to training,
validation and test sets. Each frame is represented by one ESM3 structural token per residue \cite{hayes2025esm3}. The model conditions on the sequence, the physical temperature and the current frame, together with up to three earlier frames taken 10, 20 and 30 stored frames back. Without further training, we evaluated the model on 12 mdCATH domains with no trajectory in the training set, to test transfer to unseen proteins; on Trp-cage, to test the response to physical temperature; and on L99A T4L, to test exploration of rare conformational states. 

\begin{figure*}[t]
\centering
\includegraphics[width=\linewidth]{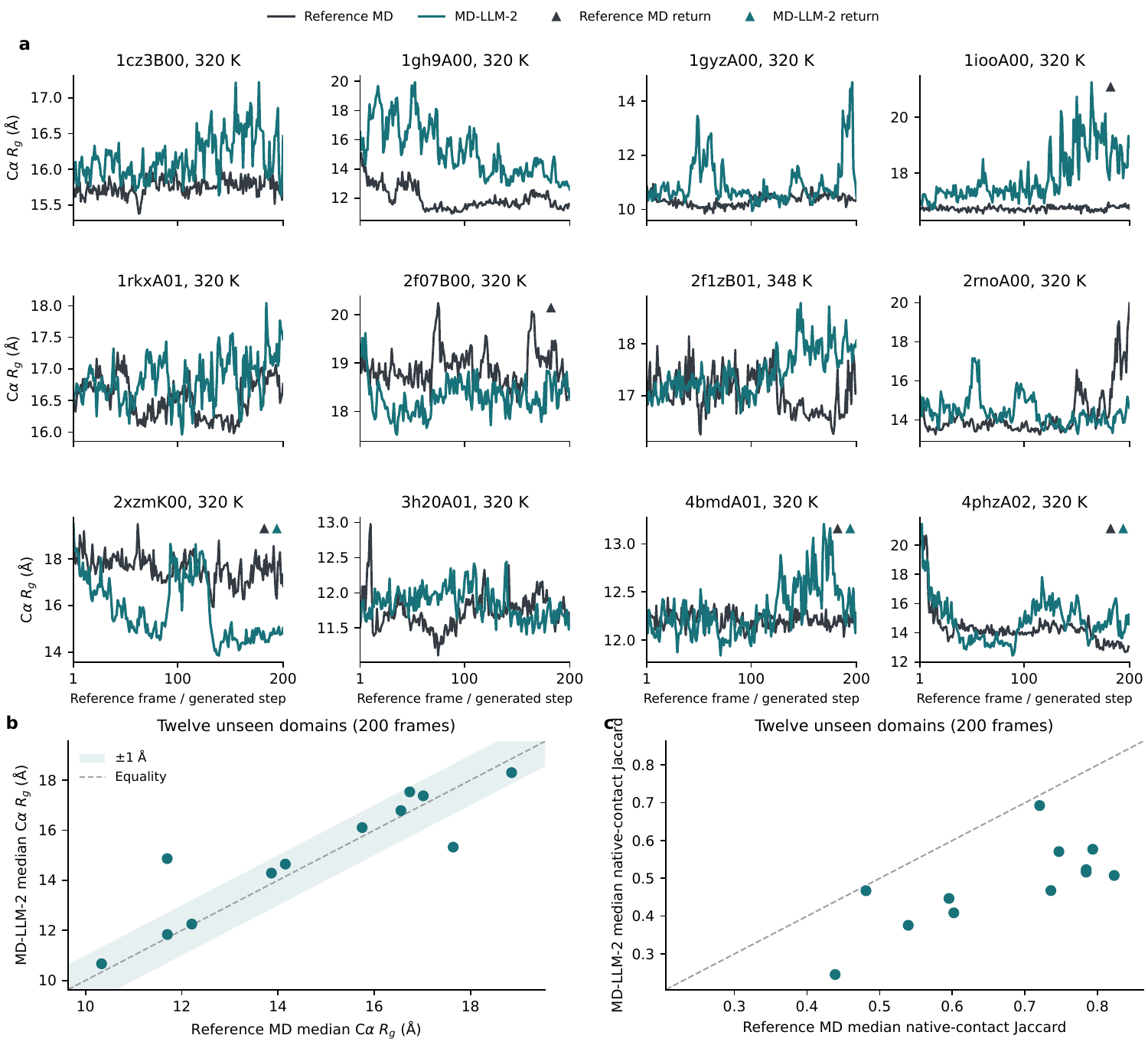}
\caption{\textbf{Generation for mdCATH domains absent from training,
compared with original MD.} (a) C$\alpha$ radius of gyration across the first 200 reference MD frames and 200 generated steps for each of the 12 domains, with the temperature of the reference trajectory. Triangles mark trajectories that depart from and return to compact conformations under the criterion of Supplementary Methods. (b, c) Domain-level medians of the radius of gyration and of the native-contact Jaccard index, generated against MD; dashed lines mark equality and the band in (b) spans $\pm1$~\AA{}. All generated trajectories use the history-conditioned model with frame-isoentropic sampling and seed 3407.}
\label{fig:mdcath}
\end{figure*}

On the 12 unseen mdCATH domains, generated trajectories reproduce the overall size of the MD ensemble (Fig.~\ref{fig:mdcath}). For each domain we generated one 200-frame trajectory with frame-isoentropic sampling from the first frame of its mdCATH trajectory, and compared it with the original. The median C$\alpha$ radius of gyration was within 1~\AA{} of the MD median in 10 of the 12 domains, with a median absolute difference of 0.39~\AA{}. The MD trajectories departed from and returned to compact conformations within 200 frames in five domains, and the generated trajectories did so in three of these and in none of the other seven.

\begin{figure*}[t]
\centering
\includegraphics[width=\linewidth]{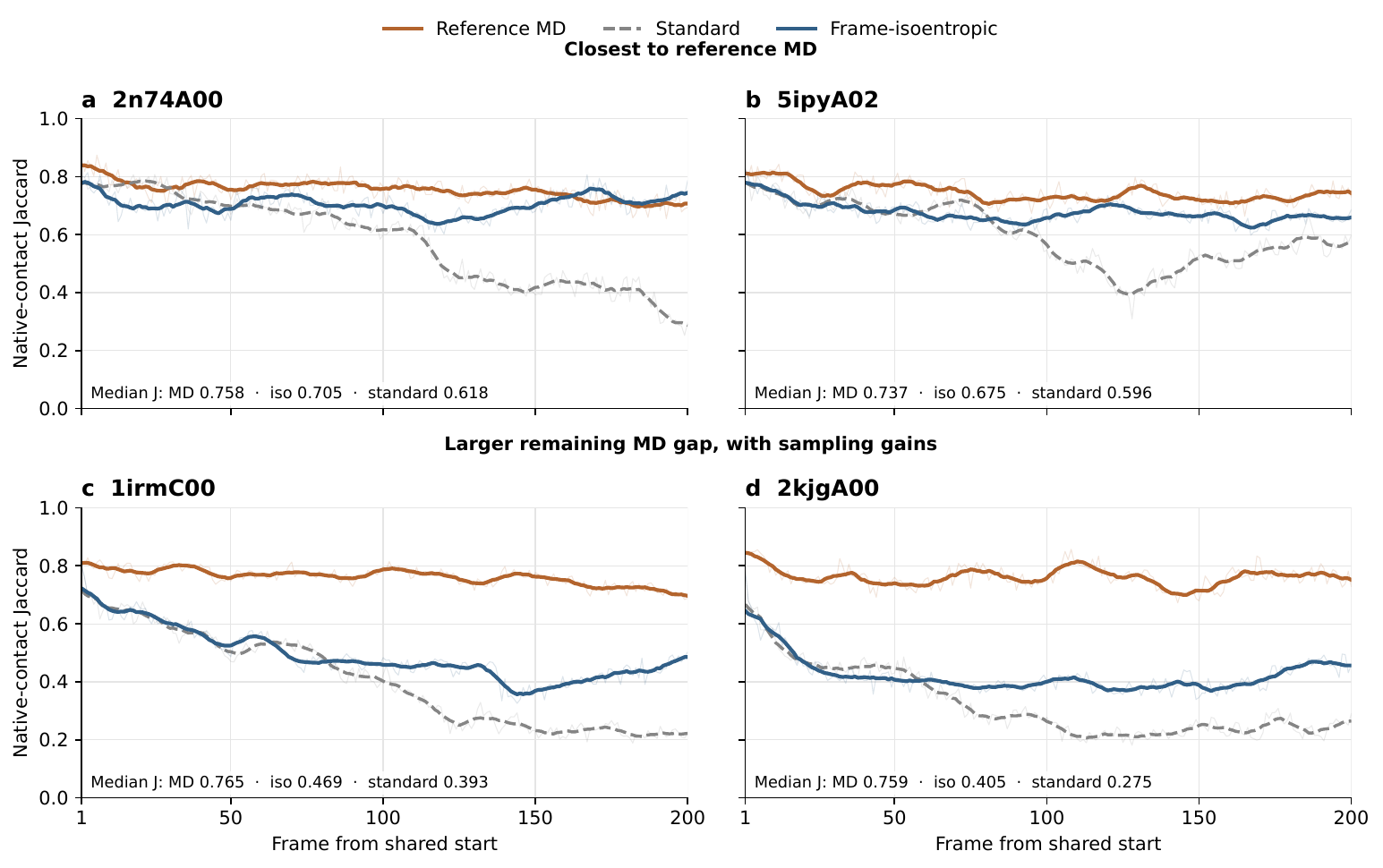}
\caption{\textbf{Frame-isoentropic sampling halts the loss of native contacts
in a model trained without history.} Four 320~K trajectories from mdCATH domains with 200-frame MD references: the two smallest (a,b) and two largest (c,d) gaps between the MD and frame-isoentropic contact medians. Standard and frame-isoentropic trajectories use the same model, trained without history, with the same starting structure, seed and one entropy probe ($M=1$); the original MD starts from the same structure and evolves independently. Faint lines show per-frame values and solid or dashed lines 11-frame moving means. Insets give the 200-frame medians. Across all ten held-out domains, frame-isoentropic sampling raises the median over standard sampling.}
\label{fig:mdcathiso}
\end{figure*}

\begin{figure*}[t]
\centering
\includegraphics[width=0.98\linewidth]{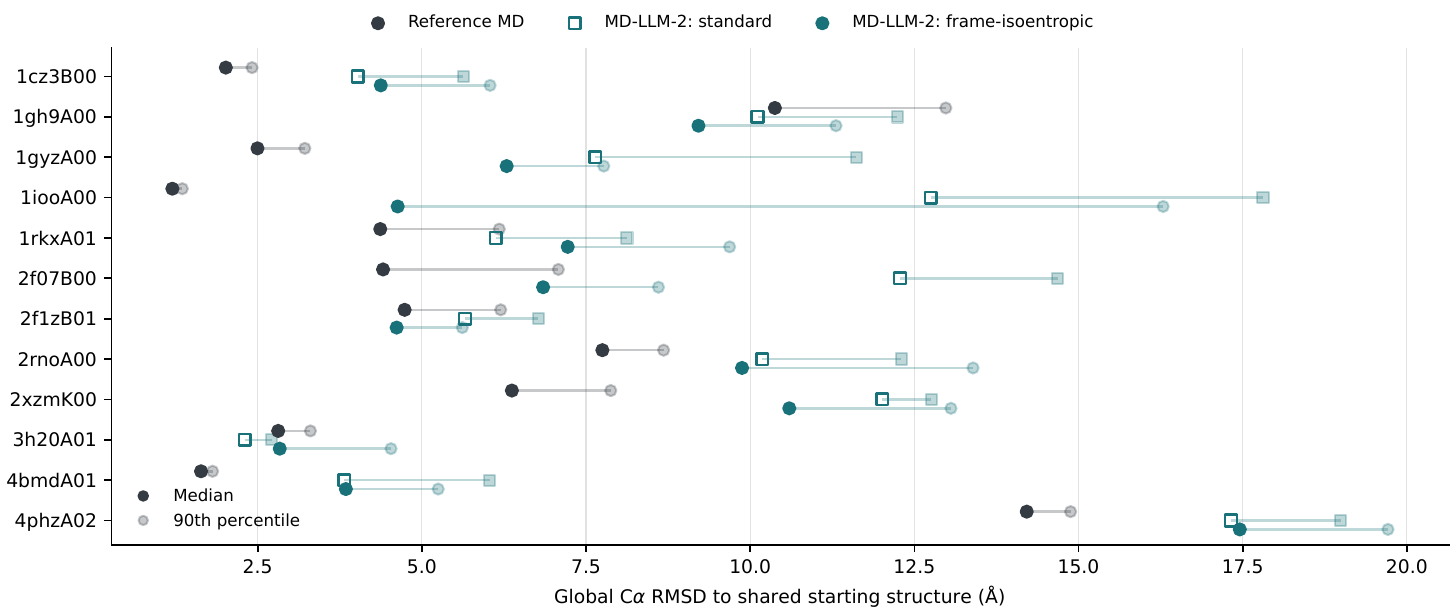}
\caption{\textbf{Standard and frame-isoentropic sampling on mdCATH domains
absent from training.} Median and 90th-percentile global C$\alpha$ RMSD to the shared starting structure over 200 frames, for the reference MD and for the history-conditioned model under standard and frame-isoentropic sampling. Both sampling rules start from the same structure with seed 3407.}
\label{fig:mdcathrmsd}
\end{figure*}

Frame-isoentropic sampling corrects drift most where it is strongest. In a model trained without history, it halted the progressive loss of native contacts seen under standard sampling: across held-out 320~K trajectories of ten training domains, generated from the same starting structure and seed under both rules, the 200-frame median Jaccard index increased in all ten, with a median gain of 0.095 (Fig.~\ref{fig:mdcathiso}). In the history-conditioned model, which already limits drift, its effect on the 12 unseen domains was concentrated in the largest departures from MD (Fig.~\ref{fig:mdcathrmsd}). The mean absolute difference between the generated and MD median C$\alpha$ RMSD from the starting structure fell from 3.6~\AA{} under standard sampling to 2.3~\AA{}, and the largest difference from 11.5 to 4.2~\AA{}, while the median difference was unchanged, at 2.3 and 2.4~\AA{}.

\begin{figure*}[t]
\centering
\includegraphics[width=0.98\linewidth]{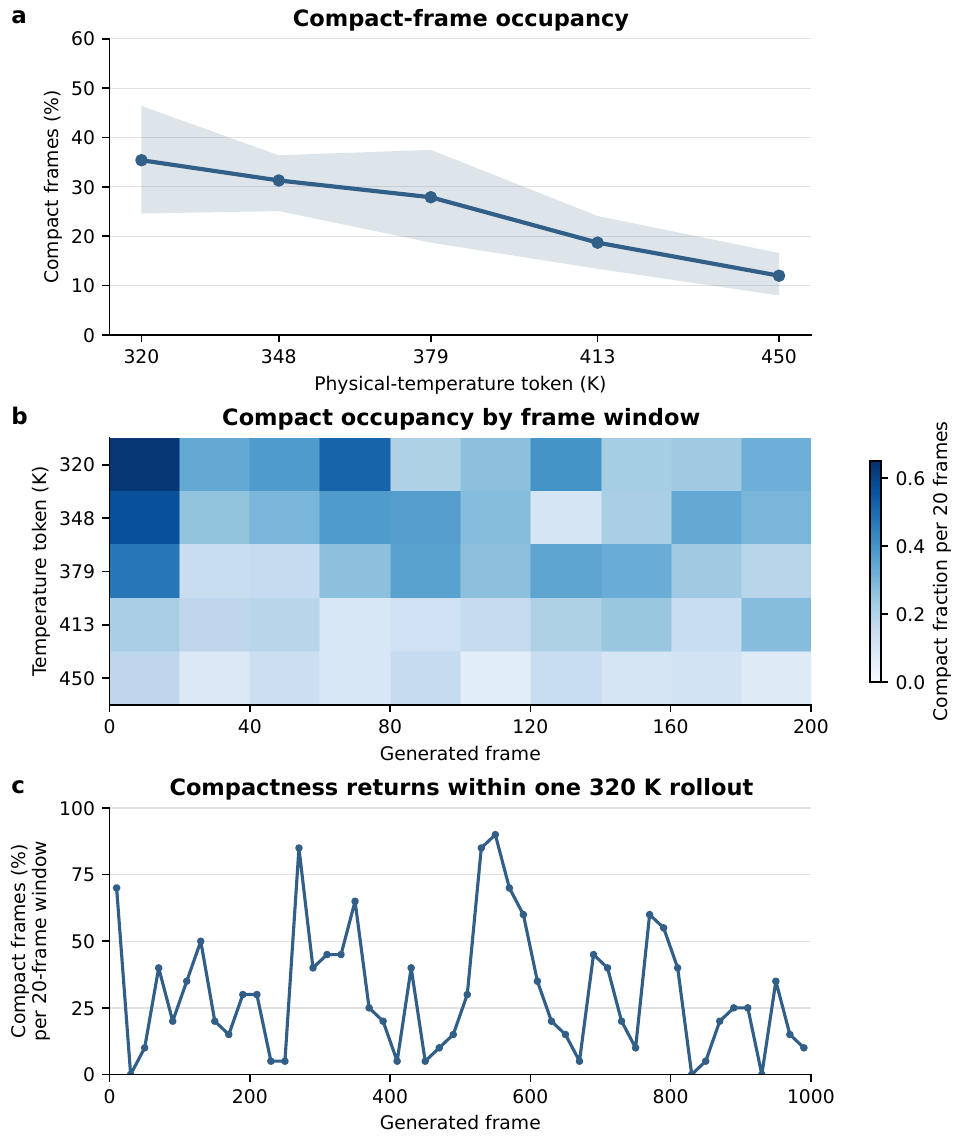}
\caption{\textbf{Temperature dependence of compact Trp-cage conformations.}
(a) Percentage of the first 200 generated frames meeting the compact criterion ($R_g\leq8.25$~\AA{} and at least 16 C$\alpha$ contacts) at each training temperature. The line pools five trajectories per temperature; shading shows 95\% bootstrap intervals over whole trajectories. (b) Compact fraction in ten consecutive 20-frame windows; each cell pools the corresponding 100 frames from five trajectories. (c) Compact fraction in consecutive 20-frame windows of one 1,000-frame trajectory at 320~K, showing repeated departures from and returns to compact conformations. All trajectories use the same model and frame-isoentropic sampling.}
\label{fig:trp}
\end{figure*}

For Trp-cage \cite{barua2008trpcage}, an external system not included in mdCATH, compaction decreases as the conditioned temperature rises. We generated five trajectories with frame-isoentropic sampling at each of the five training temperatures from 320 to 450~K, starting from the same structure, and classified a frame as compact when its C$\alpha$ radius of gyration was at most 8.25~\AA{} and it retained at least 16 C$\alpha$ contacts. Compact occupancy decreased monotonically across the five temperatures, from 35.4\% at 320~K to 12.0\% at 450~K (Fig.~\ref{fig:trp}a), and occupancy at 450~K was below that at 320~K in every 20-frame window and in each of the five paired trajectories (Fig.~\ref{fig:trp}b). At 320~K, extended 1,000-frame trajectories make repeated reversible excursions from compact conformations, and compact frames still make up 20 to 35\% of frames 501 to 1,000 in every trajectory, comparable to the occupancy over the first 200 frames (Fig.~\ref{fig:trp}c and Supplementary Fig.~\ref{fig:trp1000}).

\subsection{Molecular history and conformational transitions in T4L}
\label{sec:t4l}

L99A T4L interconverts between a ground state, E, in which Phe114 is solvent-exposed, and an excited state, B, populated at about 3\% with a lifetime of about 1~ms, in which Phe114 is buried in the cavity created by the mutation \cite{bouvignies2011excited}. Because E persists for tens of milliseconds in L99A, the E-to-B transition is challenging in MD. Vallurupalli et al.\ therefore studied the same conformational exchange in the related L99A/G113A/R119P triple mutant, in which the transition is accelerated, and described it with a 19-state Markov state model ordered by committor \cite{vallurupalli2016atomistic}. State 0 corresponds to the E-like ground state, states 8 and 9 lie in the transition region with committor values near 0.5, and the late states approach the B-like excited-state ensemble. Starting from the E-state structure (PDB 3DMV \cite{liu2008pdb3dmv}), we first examined how molecular history affects the stability of generated trajectories and then compared the generated paths with these published structures.

Conditioning on recent frames keeps the T4L scaffold intact during generation (Table~\ref{tab:t4lhistorystructure}). From the same starting structure and five seeds, MD-LLM-2 generated trajectories with a median C$\alpha$ RMSD of 3.4~\AA{} from the start when given its history of earlier frames, compared with 12.0~\AA{} when its history slots were filled with the current frame and 15.5~\AA{} for a model trained without history. Adding frame-isoentropic sampling kept the median at 3.4~\AA{} and reduced the 90th-percentile RMSD from 6.3 to 4.9~\AA{}, with improvement in four of the five trajectories.

\begin{table*}[t]
\centering
\caption{\textbf{Molecular history and frame-isoentropic sampling limit T4L
scaffold drift.} Global C$\alpha$ RMSD to the 3DMV starting structure over the first 200 generated frames of five trajectories. Entries are the median across trajectories of each trajectory's median or 90th-percentile RMSD. Standard sampling uses $\tau=1$; frame-isoentropic sampling uses a 16-probe entropy reference.}
\label{tab:t4lhistorystructure}
\small
\begin{tabular}{llll}
\toprule
Model and history & Sampling & Median (\AA{}) & 90th pct. (\AA{}) \\
\midrule
Trained without history & Standard & 15.5 & 18.4 \\ MD-LLM-2, history replaced by current frame & Standard & 12.0 & 15.0 \\ MD-LLM-2, with history & Standard & 3.4 & 6.3 \\ MD-LLM-2, with history & Frame-isoentropic & 3.4 & 4.9 \\
\bottomrule
\end{tabular}
\end{table*}

\begin{figure*}[t]
\centering
\includegraphics[width=0.98\linewidth]{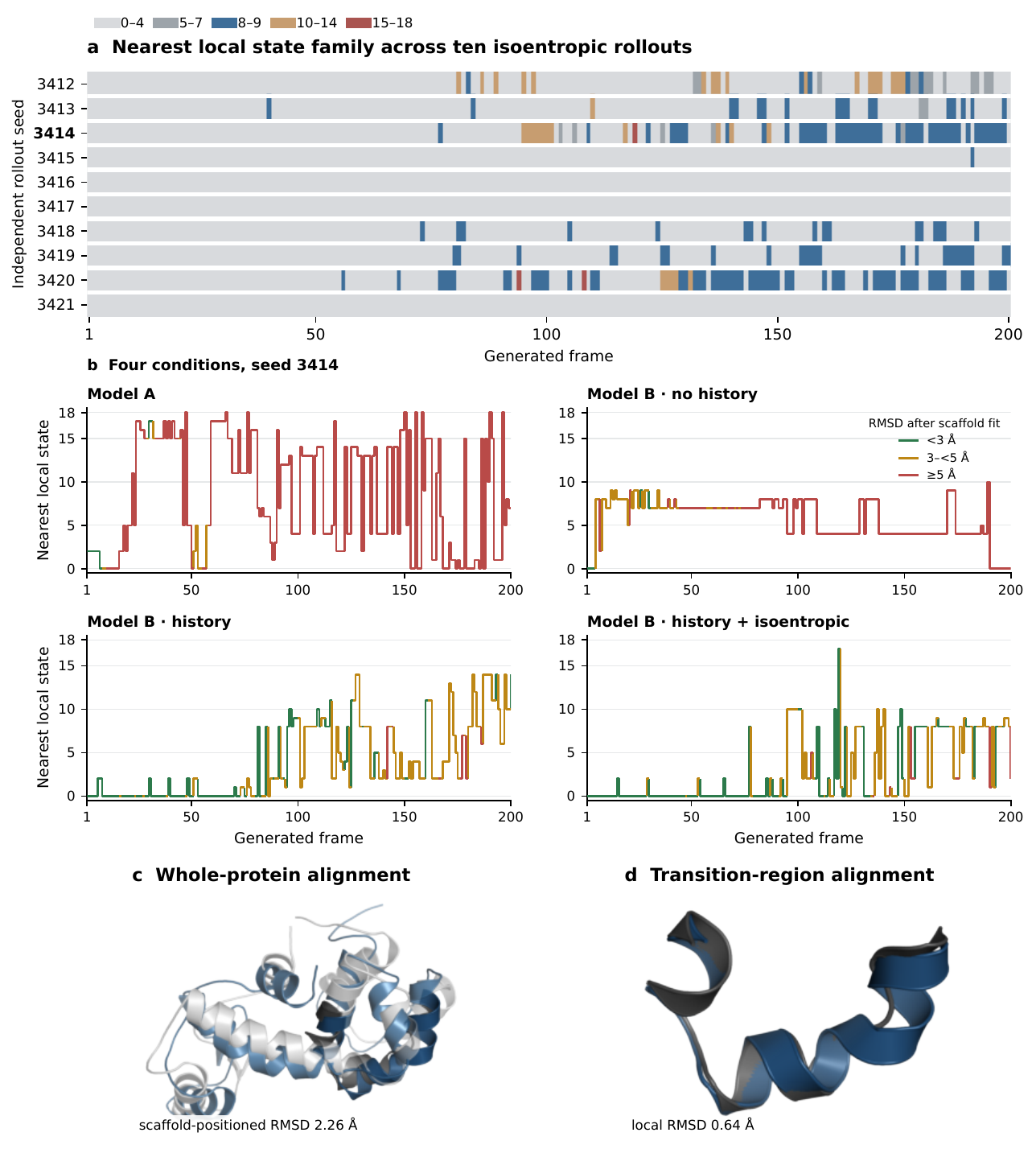}
\caption{\textbf{Generated T4L paths compared with published structural
states.} (a) Nearest published state family for each frame of ten 200-frame trajectories generated with history and frame-isoentropic sampling; states 8 and 9 form the transition region and states 15 to 18 the B family
\cite{vallurupalli2016atomistic}. Seed 3414 is highlighted. (b) Nearest state
over the first 200 frames of seed 3414 for the model trained without history, MD-LLM-2 with history replaced by the current frame, and MD-LLM-2 with history under standard or frame-isoentropic sampling. Each frame is assigned the published structure with the lowest locally fitted backbone RMSD among 95 structures; line colour gives the transition-region RMSD to that structure after global scaffold alignment (green below 3~\AA{}, amber 3 to 5~\AA{}, red 5~\AA{} or above). (c) Whole-protein overlay after scaffold alignment and (d) residues 100 to 120 after local fitting, comparing frame 170 of seed 3414 (blue) with its nearest published state-8 structure (grey). RMSDs use the 66 backbone atoms common to generated and published structures. The published ensemble represents L99A/G113A/R119P; generated trajectories use L99A.}
\label{fig:t4lpath}
\end{figure*}

Within this stable scaffold, the generated trajectories reach the transition region repeatedly (Fig.~\ref{fig:t4lpath}a). We generated ten 200-frame trajectories with history and frame-isoentropic sampling and assigned each frame to the published structure with the lowest locally fitted RMSD over the backbone of residues 100 to 120 and Leu133, among five representative structures from each of the 19 states. Because the model generates backbone coordinates only, this assignment reflects the backbone changes that accompany the transition, including the rotation of the Phe114 $\psi$ angle and the merging of helices f and g, and not the burial of the Phe114 sidechain itself. In seven of the ten trajectories, a transition-region structure (states 8 and 9) became the nearest reference, with best local RMSDs of 0.64 to 1.03~\AA{}, and the trajectories returned to E-like states (0 to 4) between visits (Supplementary Fig.~\ref{fig:t4lwalkers}). Four of these trajectories also placed the transition region within 3~\AA{} of the reference after global scaffold alignment. Two trajectories generated three frames whose nearest reference was state 17, in the B family (excited), with local RMSDs of 0.97 to 1.35~\AA{} and 1.83 to 2.64~\AA{} after global alignment; in both, these frames appeared after the trajectory had first visited the transition region.

Reaching the transition region requires more than structural motion of the same size. We compared the generated paths with torsion-angle random walks of the protein chain whose frame-to-frame C$\alpha$ displacements were drawn from the pooled displacements of the generated trajectories. At comparable global drift from the starting structure, the random walks never approached the transition-region structures as closely as the generated paths, in any of the 13 drift bins containing both groups (Supplementary Fig.~\ref{fig:t4lnull}).

\section{Discussion}

These results show that molecular trajectories can be represented as conditional distributions over tokenized paths, conditioned on molecular history and thermodynamic state. Because the model assigns an explicit probability to each path, likelihood ratios between models and the reweighting of generated paths can be evaluated directly, without estimating densities in continuous configuration space.

Of the conditioning variables, molecular history had the largest effect on the protein results. In T4L, filling the history slots with the current frame raised the median scaffold RMSD from 3.4 to 12.0~\AA{}, and a model trained without history drifted further still, to 15.5~\AA{}, so the effect does not arise only from presenting a history-trained model with unfamiliar input. The Mori--Zwanzig formalism offers one interpretation: at the stored-frame lag and in a backbone token representation, the projected dynamics retain memory that a single frame cannot supply. History may also make generation more robust to accumulated errors, since earlier frames anchor each prediction to structures that preceded any drift. A memory effect should weaken once the lag exceeds the memory time, whereas robustness to accumulated error would be expected to persist, so varying the lag would separate the two.

Frame-isoentropic sampling acts only on the token probabilities, leaves the physical temperature and coupling unchanged and does not impose detailed balance, so it corrects for exposure bias in the generative model, while its relation to a thermostat remains conceptual. With history, it acted mainly on the largest departures from MD, as in T4L, consistent with history already limiting typical drift. An early drop in native contacts in some domains was unaffected by the sampling rule and may reflect imperfect transfer or the reconstruction error of the structural tokeniser.

The present work has four main limitations. First, the protein model generates backbone coordinates only, so sidechain-dependent events, such as the burial of Phe114 in T4L, can be inferred only from their backbone signatures; an all-atom representation would allow such events to be resolved directly. Second, Hamiltonian conditioning was demonstrated only in alanine dipeptide, because mdCATH varies temperature but not the Hamiltonian; extending it to proteins will require datasets with controlled Hamiltonian variation. Third, the protein lag is fixed by the stored-frame interval of mdCATH, and the timescales of generated protein trajectories have not yet been calibrated against MD. Fourth, the kinetic test is limited to first-passage probabilities between two basins of alanine dipeptide over 100 ps. A quantitative assessment of rates and timescales, across systems of increasing complexity and including proteins, is the subject of ongoing work.

The computational cost of generating candidate trajectories that approach these conformational states is substantially lower than that of sampling them with direct MD. Unbiased MD of L99A T4L in explicit solvent (32,686 atoms) reaches 655~ns/day on one A100 with a 4-fs timestep and hydrogen mass repartitioning, and 353~ns/day at 2~fs, so sampling the millisecond timescale of the E-to-B exchange would take about 4~GPU-years. Generating the ten 200-frame trajectories that approach the published transition-region structures took 16.76 A100-hours. The comparison concerns the wall-clock cost of producing candidate conformations: generated trajectories are not calibrated in physical time, and reproducing populations and rates for proteins remains open.

More generally, MD-LLM-2 learns a conditional distribution \(q_\theta(\zframe \mid c)\), in which the molecular representation \(\zframe\) and the conditioning variables \(c\) can be chosen according to the physical problem. In the present work, this formulation is applied using different discrete representations for proteins and alanine dipeptide, with conditioning on temperature, Hamiltonian coupling and molecular history as appropriate. The same framework could in future be extended beyond trajectory generation by incorporating other sources of physical or experimental information. Experimental observables could be used to condition the generation of ensembles consistent with measurement, while mutations or binding partners could condition changes in conformational populations. Discrete representations of molecular interactions could also enable the modelling of condition-dependent binding states. These extensions would require suitable representations and training data, but not a change in the underlying probabilistic formulation.

\section{Methods}

\paragraph{Training objective and path probabilities.}
We denote the map from coordinates to tokens by $E$ and the decoder from tokens to coordinates by $D$. For a protein with $N$ residues, $E(\xframe_t)$ gives one ESM3 structural token per residue; the alanine tokenizers are described in Supplementary Methods. All models are trained with teacher forcing: when each next token is scored, the preceding tokens of the target frame are taken from the data, and context tokens receive no loss. For the protein model, let $\widetilde q_\theta$ be the language model's distribution over its full vocabulary, and let $\mathcal T_{i,t}$ contain the residue and padding positions of the target frame in training example $(i,t)$. Up to minibatch normalization, the training objective is
\begin{equation}
\begin{aligned}
\mathcal L_{\rm protein}(\theta)
&=-\sum_{i,t}\sum_{r\in\mathcal T_{i,t}}\\ &\quad\log\widetilde q_\theta
\bigl(y^{(i)}_{t+1,r}\mid H^{(i)}_t,c_i,
y^{(i)}_{t+1,<r}\bigr),
\end{aligned}
\label{eq:training}
\end{equation}
where $y$ is a structural token or the padding token, and context, sequence and separator positions are masked. The alanine models instead use output heads specific to each codebook, with cross-entropy only on their target code tokens, conditioned on $\lambda$ or on the preceding trajectory as appropriate.

Because the language model's vocabulary also contains text and control tokens, at protein inference and in likelihood analyses we restrict the logits to the 4096 valid ESM3 structural codes $\mathcal V$ and renormalize, $q^{\rm tr}_{\theta,1}(k\mid\cdot)=\widetilde q_\theta(k\mid\cdot)/
\sum_{j\in\mathcal V}\widetilde q_\theta(j\mid\cdot)$ for $k\in\mathcal V$,
where the subscript 1 denotes sampling temperature $\tau=1$. A positive sampling temperature $\tau$ rescales these logits as in Eq.~\ref{eq:entropy}. If $\pi$ is a fixed or causal rule for choosing $\tau_s$, the probability of a token path conditional on the initial frame is
\begin{equation}
\begin{aligned}
Q_{\theta,\pi}(\zframe_{1:L}\mid\zframe_0,c) &=\prod_{s=1}^{L}\prod_{r=1}^{N}\\ &\quad q^{\rm tr}_{\theta,\tau_s} (z_{s,r}\mid H_{s-1},c,z_{s,<r}),\\
\tau_s&=\pi(\zframe_{0:s-1},c).
\end{aligned}
\label{eq:pathlaw}
\end{equation}
Each factor is normalized over the valid structural codes, so Eq.~\ref{eq:pathlaw} defines a normalized distribution over token paths whenever the sampling rule is causal. This distribution is conditional on the starting frame and the sampling rule, including any probes used to set the entropy target.

\paragraph{Model instances and training data.}
All models share the within-frame autoregressive factorisation of Eq.~\ref{eq:ar} but differ in representation and conditioning. Because mdCATH contains temperature replicas but no controlled $\lambda$ series, the alanine models are trained on a separate Hamiltonian-coupling dataset. The alanine equilibrium model defines $q^{\rm eq}_\theta(\zframe\mid\lambda)$ and emits two codes per frame; ten independently seeded instances form the population ensemble, and a further instance trained on the same couplings is used for the $\lambda=0$ sampling test. The alanine transition model emits one $\phi/\psi$ grid token per step and conditions on the preceding frames. The protein model emits one ESM3 structural token per residue, drawn from 4096 valid codes, which the ESM3 structure encoder and decoder (\texttt{esm3\_sm\_open\_v1}) map from and to backbone N, C$\alpha$ and C coordinates \cite{hayes2025esm3}. All models use a Qwen2.5-14B backbone adapted with rank-64 low-rank updates on the attention and feed-forward projections \cite{yang2024qwen25,hu2022lora}.

In the protein model, the sequence and discrete tokens for physical temperature and lag precede the structural context in the prompt. The five temperature tokens are 320, 348, 379, 413 and 450~K, and all evaluations here use the lag token for one stored frame (1~ns in mdCATH). Training uses replica 0 of the mdCATH trajectories of domains with at most 200 residues, at all five temperatures. Individual trajectories were assigned at random to training, validation and test sets in the proportions 70:20:10 before temporal contexts were built, giving 7,275,485 training, 2,084,666 validation and 1,040,992 test transition windows; a domain can therefore contribute trajectories at different temperatures to different sets. The validation set was used only to monitor the loss. The history consists of the frames at indices $t-30$, $t-20$, $t-10$ and $t$, in that order; when an older index falls before the start of the trajectory, the earliest available frame is included once. No context window crosses a trajectory or split boundary. 

The 12 mdCATH domains used to test transfer have no trajectory in the training set, and Trp-cage and L99A T4L were not used for training. Unless stated otherwise, protein comparisons use the first 200 generated frames for every system and sampling rule, and comparisons with mdCATH use the first 200 frames of the reference trajectory. For each unseen domain we use its 320~K trajectory, except 2f1zB01, whose 320~K trajectory has fewer than 201 saved frames and for which the 348~K trajectory is used instead. 

\paragraph{Alanine dipeptide.}
The system is 22-atom ACE--ALA--NME in implicit solvent, modelled with AMBER ff96 and OBC2 at 300~K \cite{kollman1997ff96,onufriev2004obc}. The dimensionless coupling $\lambda$ scales the torsional energy,
\begin{equation}
U_\lambda(x)=U_0(x)+\lambda U_1(x),\qquad U_1(x)=\sum_{f\in\mathcal F_{\rm torsion}}U_f(x),
\label{eq:hamiltonian}
\end{equation}
where $\mathcal F_{\rm torsion}$ comprises all proper and improper torsion terms of the OpenMM \texttt{PeriodicTorsionForce} \cite{eastman2017openmm} and $U_0$ contains all other terms. Thus $\lambda=0$ removes the torsional terms and $\lambda=1$ restores the original force field, while the temperature stays at 300~K. The integrator and tokenisers are described in Supplementary Methods.

Training data for the equilibrium model came from four Hamiltonian replica-exchange (HREX) chains \cite{fukunishi2002hrex} over the training couplings $\lambda=0.8$, 1.0, 1.2 and 1.3, each with 100~ps of burn-in and 500~ps of production per state. Two chains were used for training and for MBAR, and two were held out for validation and tokenizer testing. The reference populations at unseen couplings came from a separate eight-chain HREX ladder spanning $\lambda=0$--$2$ in steps of 0.1, with 100~ps of burn-in and 400~ps of production per state, generated after the model predictions had been fixed; each reference population is the mean of the eight chain fractions computed from MD coordinates. The $\lambda=0$ reference uses four independent fixed-$\lambda$ chains.

Two basins are defined by the nearer of two circular $\psi$ centres, 2.7489 (basin 0) and $-0.9163$~rad (basin 1). The equilibrium model emits two codes, $a$ and $b$, from 128-entry codebooks. Ten instances with different random seeds were trained for 1,500 optimizer updates with identical hyperparameters on individual training frames, and the sampling temperature $\tau=0.7$ was selected on validation data before the reference was generated. Writing $b_{\rm eq}(D(a,b))\in\{0,1\}$ for the basin of a decoded code pair, each model's population is obtained by enumerating all $128^2$ pairs:
\begin{equation}
\begin{aligned}
\widehat p^{(m)}_1(\lambda;\tau)
&=\sum_{a=1}^{128}\sum_{b=1}^{128} q^{\rm eq,(m)}_{\theta,\tau}(a\mid\lambda)\\ &\quad\times q^{\rm eq,(m)}_{\theta,\tau}(b\mid a,\lambda)
\,\mathbf 1\{b_{\rm eq}(D(a,b))=1\}.
\end{aligned}
\label{eq:population}
\end{equation}
The ensemble prediction averages the ten models with equal weight, and the free-energy difference is $\Delta F_{0\to1}/(k_{\rm B}T)=-\log[p_1/(1-p_1)]$. The population error is the mean absolute difference from the reference over the four unseen couplings, and reference intervals resample complete chains. The $\lambda=0$ test uses a further equilibrium model trained on the same couplings, with frame-isoentropic sampling as described in Supplementary Methods.

For MBAR \cite{shirts2008mbar}, the reduced energy of stored configuration $x_n$ at coupling $\lambda_j$ is $u_j(x_n)=[U_0(x_n)+\lambda_jU_1(x_n)]/(k_{\rm B}T)$. Writing $N_j$ for the number of training samples from state $j$ and $f_j$ for the dimensionless free energies, we solve
\begin{equation}
f_i=-\log\!\left[\sum_n
\frac{e^{-u_i(x_n)}}{\sum_j N_j e^{f_j-u_j(x_n)}}\right],
\qquad f_0=0,
\label{eq:mbarfixedpoint}
\end{equation}
and estimate the target population with the weights
\begin{align}
v_n(\lambda_*)&=\frac{e^{-u_{\lambda_*}(x_n)}} {\sum_j N_j e^{f_j-u_j(x_n)}},\nonumber\\ w_n(\lambda_*)&=\frac{v_n(\lambda_*)}{\sum_m v_m(\lambda_*)},\nonumber\\
\widehat p^{\rm MBAR}_1(\lambda_*)&=
\sum_n w_n(\lambda_*)\mathbf 1\{b_{\rm eq}(x_n)=1\}.
\label{eq:mbar}
\end{align}
The sums use the same two training chains as the model and no reference configurations. The effective sample size $N_{\rm eff}=1/\sum_n w_n(\lambda_*)^2$ measures the overlap with the target before accounting for temporal correlation.

The transition model tokenises each $\phi$ and $\psi$ angle into 24 equal periodic bins on $[-\pi,\pi)$, giving one of $24^2=576$ states per frame. It is trained on 20-ps fixed-$\lambda$ MD trajectories at $\lambda=1.0$, 1.3 and 2.0, with frames saved every 0.2~ps, conditions on the preceding frame tokens, and receives $\lambda$ through a learned continuous projection added to the coupling-token embedding. It samples at $\tau=0.85$, selected on training-state validation. A 100-ps path is generated as five successive 20-ps blocks of 100 steps, each starting from the final token of the previous block at the same $\lambda$, so the reported paths compose the learned 20-ps transition kernel. For initial core $A$ and opposite core $B$, each defined as the $\psi$ values within $\pi/3$ of a basin centre, let $b_{\rm core}$ return the core label, or $-1$ outside both cores. The cumulative first-passage probability at horizon $h$ is
\begin{equation}
\begin{aligned}
T_{A\to B}&=\delta t\,\inf\{s\geq0:b_{\rm core}(\zframe_s)=B\},\\ P_{A\to B}(h)&=\Pr\!\left[T_{A\to B}\leq h
\mid b_{\rm core}(\zframe_0)=A\right],
\end{aligned}
\label{eq:firstpassage}
\end{equation}
with $\delta t=0.2$~ps and $T_{A\to B}=\infty$ if the opposite core is never reached; frames outside both cores are neutral. At each held-out coupling, direct fixed-$\lambda$ MD provides 64 independent 100-ps paths per starting core, and the model generates 256 per core from the same starting pool. The error is the mean of $|\widehat P_{A\to B}^{\rm method}(h)-\widehat P_{A\to B}^{\rm MD}(h)|$ over both couplings, both directions and $h\in\{20,50,100\}$~ps. Its 95\% interval comes from a 20,000-draw bootstrap that resamples complete paths within each coupling and starting core, keeping horizons paired within paths and holding the fitted MSM fixed.

The MSM comparator \cite{prinz2011msm} is fitted only to the training trajectories. At each training coupling $\lambda_j$, transition counts $C^{(j)}_{uv}$ at a candidate lag are symmetrized, and on the union of observed edges we fit
\begin{align}
\log\!\left[C^{(j)}_{uv}+C^{(j)}_{vu}+\epsilon\right]
&\approx a_{uv}+(\lambda_j-1)b_{uv},\nonumber\\ F_{uv}(\lambda)&=F_{vu}(\lambda) =e^{a_{uv}+(\lambda-1)b_{uv}},\nonumber\\ P_{uv}(\lambda)&=\frac{F_{uv}(\lambda)}{\sum_w F_{uw}(\lambda)}.
\label{eq:msm}
\end{align}
States without support have self-loops, and the symmetric flux makes $P$ reversible with respect to its row-sum stationary distribution. A lag of 1~ps and pseudocount $\epsilon=1$ were selected by first-passage error on training validation data; the reference MD paths were used only for final scoring.

\paragraph{Frame-isoentropic sampling.}
The physical temperature $T$ is a conditioning input of the model, whereas the sampling temperature $\tau$ only rescales the token probabilities at inference. For the unscaled logits $\ell_{s,r,k}$ over valid codes, recorded along the realised tokens of generated frame $s$, we define the token probabilities and their mean categorical entropy
\begin{equation}
\begin{aligned}
p_{s,r}^{\tau}(k) &=\frac{\exp(\ell_{s,r,k}/\tau)} {\sum_{j\in\mathcal V}\exp(\ell_{s,r,j}/\tau)},\\ h_s(\tau)&=\frac{1}{N}\sum_{r=1}^{N}
\mathcal H\bigl(p_{s,r}^{\tau}\bigr),
\end{aligned}
\label{eq:entropy}
\end{equation}
where $\mathcal H(p)=-\sum_{k\in\mathcal V}p(k)\log p(k)$. Generation starts at $\tau_1=1$, and the reference entropy is set from $M$ independent probes of the first transition from the initial frame, $h_\star=M^{-1}\sum_{m=1}^{M} h_1^{(m)}(1)$, with $M=16$ for all protein results except the comparison with the model trained without history, which used $M=1$. After frame $s$ is generated, the controller evaluates its recorded logits at candidate temperatures and solves $h_s(\tau_{s+1})=h_\star$ by bracketing and bisection, which is well posed because $h_s$ increases monotonically with $\tau$ (Supplementary Methods). The same $\tau_{s+1}$ is used for every token of the next frame. The update is therefore delayed by one frame, uses no logits from the frame being generated, and leaves the physical temperature and coupling unchanged. The controller for the alanine equilibrium model, which acts on its two-code distribution, is described in Supplementary Methods.

\paragraph{T4L history comparison.}
Both protein models condition on the sequence, lag, physical temperature and current frame; the history-conditioned model also receives up to three earlier frames, and the model trained without history does not. Four configurations were compared: the model trained without history; the history-conditioned model with its history slots filled with the current frame; and the history-conditioned model with the chronological history of its own generated frames, under either standard sampling ($\tau=1$) or frame-isoentropic sampling ($M=16$). Early in a trajectory, when fewer than 30 frames have been generated, the history contains the starting frame following the rule above. All four configurations were run from chain A of 3DMV \cite{liu2008pdb3dmv} with the 320~K temperature token and five seeds (3407--3411), and evaluated over the first 200 generated frames. At each frame, the C$\alpha$ RMSD to 3DMV is computed after optimal global C$\alpha$ alignment, and Table~\ref{tab:t4lhistorystructure} reports the median across seeds of each trajectory's median and 90th-percentile RMSD.

\paragraph{Protein structural analyses.}
For C$\alpha$ positions $r_1,\ldots,r_N$ with centroid $\bar r$, the radius of gyration is $R_g=[N^{-1}\sum_i\|r_i-\bar r\|_2^2]^{1/2}$. Native contacts are the residue pairs $(i,j)$ with $|i-j|\geq3$ and C$\alpha$ distance below 8~\AA{}, and the native-contact Jaccard index of a frame $x$ is $J(x,x_0)=|C(x)\cap C(x_0)|/|C(x)\cup C(x_0)|$, where $C(x)$ is its contact set and $x_0$ the starting frame shared by generated and reference trajectories. Generated and reference trajectories evolve independently after this common start, so frames at the same index are not paired configurations; one trajectory is generated per unseen domain, and the comparison with the model trained without history uses the mean over three generated trajectories at each frame. For Trp-cage, the K8A variant starting from the first frame of a 2JOF-derived trajectory \cite{barua2008trpcage,lindorff2011fastfolding}, a frame is compact when $R_g\leq8.25$~\AA{} and it has at least 16 C$\alpha$ contacts within 8~\AA{} at a sequence separation of at least four. For T4L, each frame is assigned to the nearest of 95 published structures, five from each of the 19 states \cite{vallurupalli2016atomistic}, by RMSD after local fitting on the 66 backbone atoms shared by generated and published structures: N, C$\alpha$ and C of residues 100 to 120 and of Leu133. The transition-region RMSD to the assigned structure is then computed after global scaffold alignment
\cite{kabsch1976rotation}. The published ensemble represents
L99A/G113A/R119P, whose states 15 to 18 correspond structurally to the L99A excited state \cite{bouvignies2011excited}. The torsion-angle random walks perturb the backbone $\phi$ and $\psi$ angles of the chain starting from 3DMV, with frame-to-frame C$\alpha$ displacements drawn from the pooled displacements of the ten generated trajectories (Supplementary Methods). Independent trajectories are the units of replication throughout; further details are given in Supplementary Methods.

\clearpage
\appendix
\setcounter{figure}{0}
\renewcommand{\thefigure}{S\arabic{figure}}
\setcounter{equation}{0}
\renewcommand{\theequation}{S\arabic{equation}}
\setcounter{table}{0}
\renewcommand{\thetable}{S\arabic{table}}

\section{Supplementary Results}

\subsection{Predictive entropy tracks structural drift}

The model's predictive entropy rises as generated trajectories drift from the starting structure, which motivates its use as the control signal in frame-isoentropic sampling. We scored five 200-frame T4L trajectories, generated by the history-conditioned model with standard sampling, by passing each generated frame back through the same model with teacher forcing and computing its mean categorical entropy $H^{\rm cat}_t$ and mean token negative log-likelihood $L_t$ (Eq.~\ref{eq:nll}). Within each trajectory, the entropy increased with the global C$\alpha$ RMSD from the starting structure (median Spearman $\rho=0.855$), as did the negative log-likelihood (median $\rho=0.726$). For tokens sampled at $\tau=1$, the expected negative log-likelihood equals the entropy, and after subtracting the entropy, the remaining association with drift was weak (median $\rho=0.102$).

\subsection{Population calibration with two MD anchors}

\begin{figure}[t]
\centering
\includegraphics[width=0.98\columnwidth]{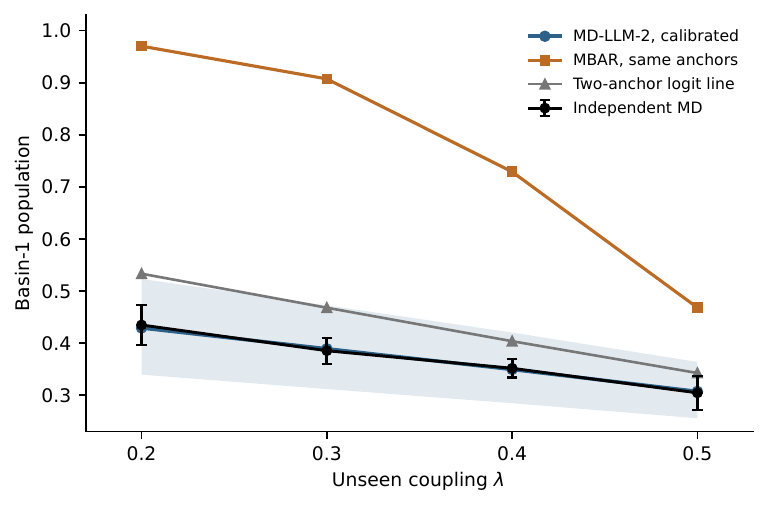}
\caption{\textbf{Population calibration with two MD anchors.} Basin-1
populations at four unseen couplings from independent MD, the calibrated MD-LLM-2 ensemble, calibrated MBAR and a straight line through the two anchor populations in log-odds space. Black error bars are 95\% intervals over the eight reference chains; blue shading is the 95\% bootstrap interval for the calibrated ensemble over anchor chains. All calibrated predictors use the same MD populations at $\lambda=0.9$ and 1.1.}
\label{fig:alacal}
\end{figure}

Two independent MD populations at couplings inside the training range sharpen the absolute population estimates. Fixed-$\lambda$ MD gave basin-1 populations of 0.1546 at $\lambda=0.9$ and 0.0977 at $\lambda=1.1$, from 64 and 32 independent chains, respectively. For each of the ten models, we fitted a two-parameter affine map of its basin-1 log odds to these anchors (Eq.~\ref{eq:calibration}) and averaged the calibrated predictions; MBAR received the same anchors and calibration. At the four unseen couplings, the calibrated ensemble reached a mean absolute error of 0.0036, compared with 0.0918 before calibration and 0.3996 for calibrated MBAR, and every calibrated model was more accurate than calibrated MBAR (Fig.~\ref{fig:alacal}). In a paired bootstrap over chains, the error of calibrated MBAR exceeded that of the calibrated ensemble by 0.3960 (95\% interval 0.2097 to 0.4074). A straight line through the two anchor log odds, using no model, reached an error of 0.0678. The calibrated ensemble is more accurate at the measured anchors, but once the uncertainty of the anchor populations is included, the difference between the two is not resolved (95\% interval $-0.0498$ to 0.0865).

\subsection{Kinetic errors by coupling and direction}

MD-LLM-2 has the lower first-passage error at each held-out coupling (Table~\ref{tab:kineticsplit}). Its mean absolute error in cumulative first-passage probability, averaged over 20, 50 and 100~ps, was 0.0794 at $\lambda=1.15$ and 0.0827 at $\lambda=0.8$, compared with 0.1666 and 0.1469 for the MSM. Each coupling on its own leaves the difference unresolved, with 95\% path-bootstrap intervals of $-0.009$ to 0.176 at $\lambda=1.15$ and $-0.041$ to 0.138 at $\lambda=0.8$, although MD-LLM-2 had the lower error in 96.1\% and 88.1\% of bootstrap draws, respectively. The primary comparison pooling both couplings excludes zero (0.005 to 0.135). At both couplings, the largest difference arose for transitions from basin 0 to basin 1, which the MSM predicted to occur too quickly.

\begin{table}[t]
\centering
\caption{\textbf{First-passage errors by coupling and direction.} Mean
absolute error in cumulative first-passage probability, averaged over the 20, 50 and 100-ps horizons. Basin 0 is centred at $\psi\approx2.75$~rad and basin 1 at $\psi\approx-0.92$~rad; lower is better.}
\label{tab:kineticsplit}
\small
\begin{tabular}{llll}
\toprule
$\lambda$ & Direction & MD-LLM-2 & MSM \\
\midrule
0.8  & $0\to1$ & 0.0820 & 0.2295 \\ 0.8  & $1\to0$ & 0.0833 & 0.0643 \\ 0.8  & Both    & 0.0827 & 0.1469 \\
\midrule
1.15 & $0\to1$ & 0.0911 & 0.3221 \\ 1.15 & $1\to0$ & 0.0677 & 0.0112 \\ 1.15 & Both    & 0.0794 & 0.1666 \\
\bottomrule
\end{tabular}
\end{table}

\subsection{Extended Trp-cage trajectories}

\begin{figure}[t]
\centering
\includegraphics[width=0.98\columnwidth]{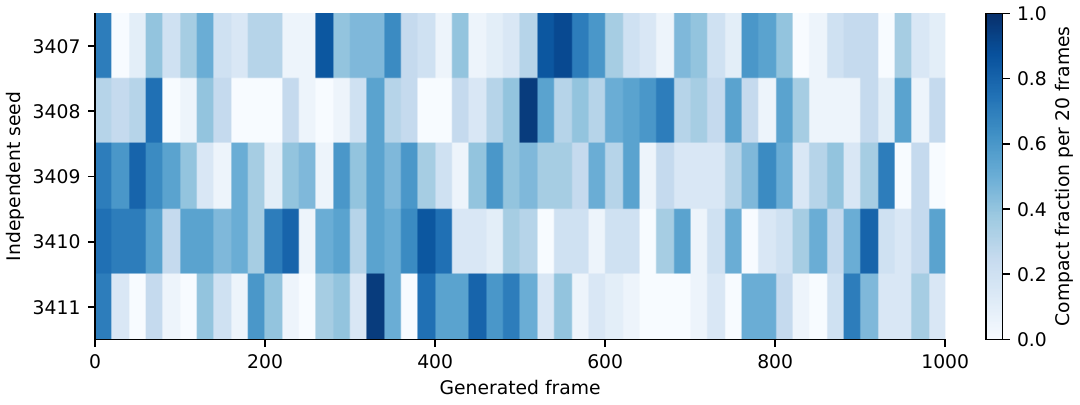}
\caption{\textbf{Recurrent compact conformations in extended Trp-cage
trajectories.} Five independent 1,000-frame trajectories at 320~K, one per row (seeds 3407--3411). Each cell gives the fraction of 20 consecutive generated frames meeting the compact criterion of Fig.~\ref{fig:trp}; darker blue indicates more compact frames. Windows rich in compact frames recur after windows with low compact occupancy throughout all five trajectories.}
\label{fig:trp1000}
\end{figure}

Compact Trp-cage conformations persist and recur over 1,000 generated frames. The five 320~K trajectories of the temperature sweep were generated for 1,000 frames, and every frame was scored with the compact criterion of Fig.~\ref{fig:trp}. Compact frames made up 20.2 to 35.4\% of frames 501 to 1,000 in every trajectory, comparable to the occupancy over the first 200 frames, and each trajectory departed from and returned to compact conformations several times (Supplementary Fig.~\ref{fig:trp1000}).

\subsection{T4L displacement-matched control}

\begin{figure}[t]
\centering
\includegraphics[width=0.98\columnwidth]{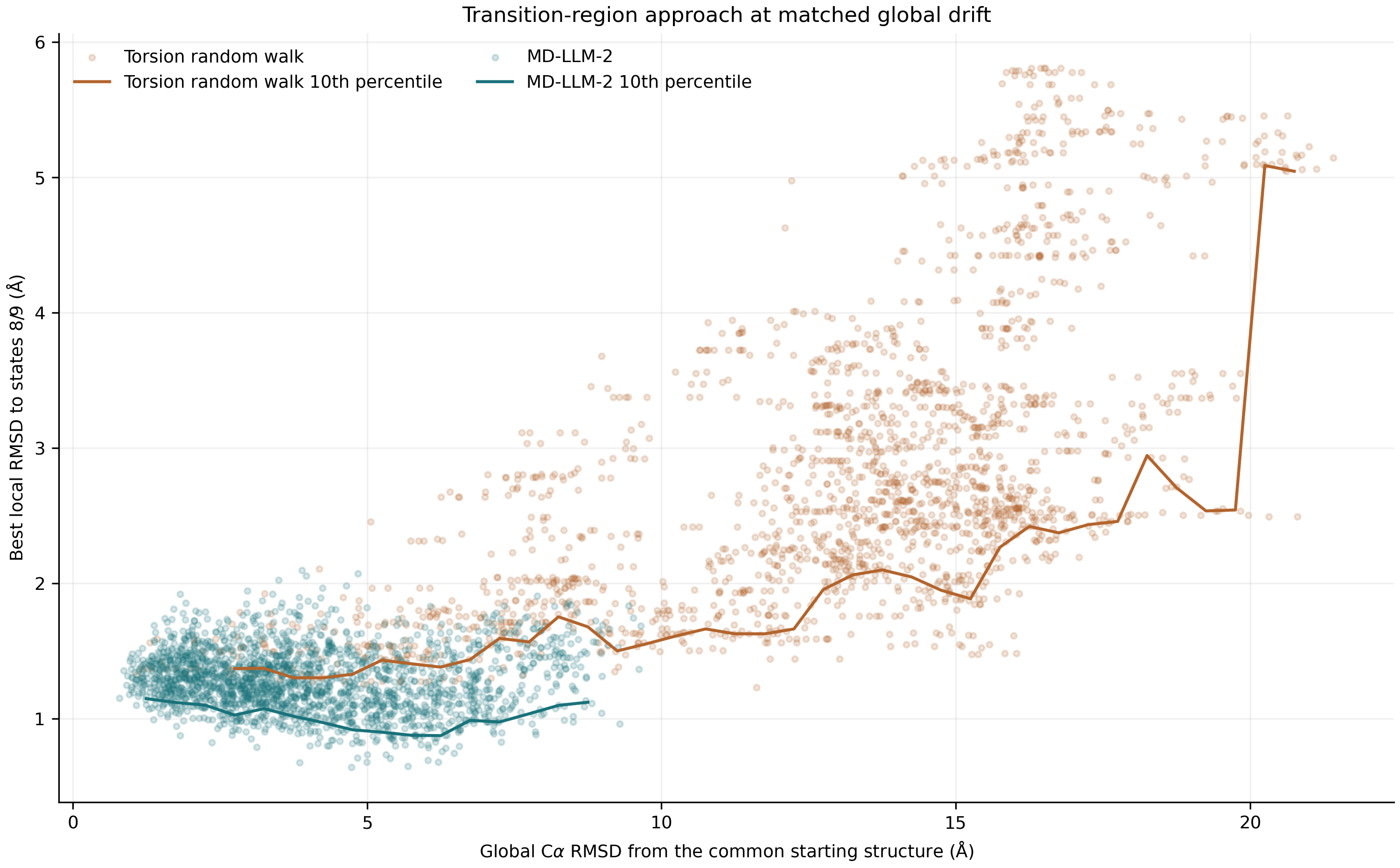}
\caption{\textbf{Generated T4L paths approach transition-region structures
more closely than displacement-matched random walks.} Blue, the ten MD-LLM-2 trajectories of Fig.~\ref{fig:t4lpath}; orange, ten torsion-angle random walks whose frame-to-frame C$\alpha$ displacements are drawn from the pooled displacements of those trajectories. All paths contain 200 frames and start from the same structure. Each point is one frame: its vertical position is the locally fitted RMSD to the transition-region states 8 and 9, and its horizontal position the global C$\alpha$ RMSD from the start. Lines show the 10th percentile of the local RMSD within bins of global RMSD; the blue line lies below the orange line in all 13 bins containing both groups.}
\label{fig:t4lnull}
\end{figure}

The generated T4L paths approach the transition-region structures more closely than random structural motion of the same size. Each control is a torsion-angle random walk of the protein chain that starts from the same structure as the generated trajectories, with frame-to-frame C$\alpha$ displacements drawn from their pooled distribution (Supplementary Methods). To compare frames at a similar distance from the start, we binned all frames by global C$\alpha$ RMSD from the starting structure and, within each bin, compared the 10th percentile of the locally fitted RMSD to the published states 8 and 9. The generated paths were closer to these structures in all 13 bins containing frames from both groups (Supplementary Fig.~\ref{fig:t4lnull}).

\begin{figure*}[t]
\centering
\includegraphics[width=0.98\linewidth]{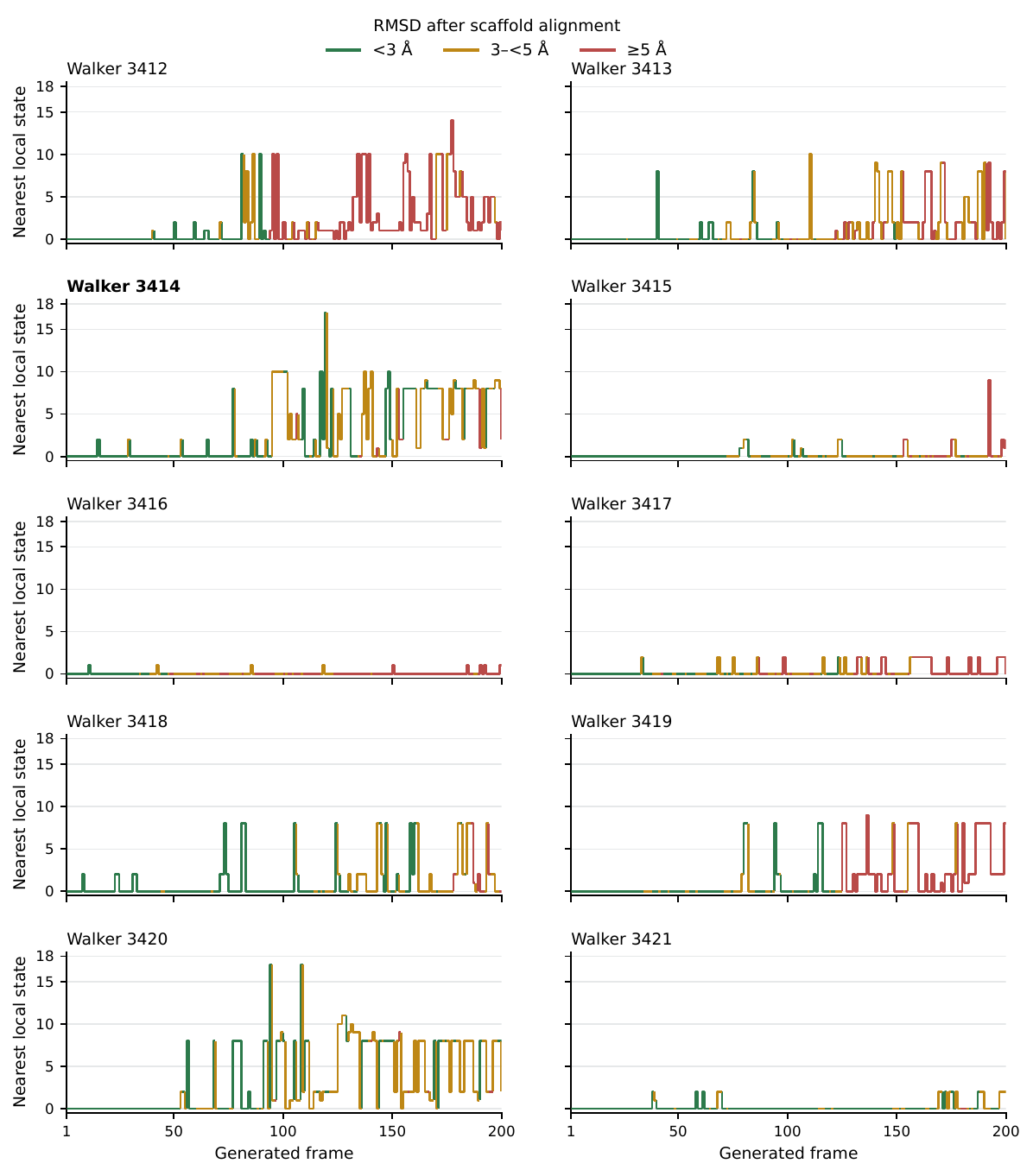}
\caption{\textbf{Published-state timelines for ten T4L trajectories.} Each
panel shows one 200-frame trajectory generated by MD-LLM-2 with history and frame-isoentropic sampling ($M=16$), seeds 3412--3421. At every frame, the plotted state (0 to 18) is that of the published structure with the lowest locally fitted RMSD among 95 structures \cite{vallurupalli2016atomistic}; states 8 and 9 form the transition region and states 15 to 18 the B family. State 17 is the nearest reference for three frames in trajectories 3414 and 3420. Line colour gives the transition-region RMSD to the same structure after global C$\alpha$ scaffold alignment: green below 3~\AA{}, amber 3 to 5~\AA{}, and red 5~\AA{} or above. The state assignment and colours follow Fig.~\ref{fig:t4lpath}b, where trajectory 3414 also appears.}
\label{fig:t4lwalkers}
\end{figure*}

\section{Supplementary Methods}

\paragraph{Alanine integration and tokenisers.}
All alanine simulations use the same overdamped Langevin update
\cite{ermak1978brownian},
\begin{equation}
\begin{aligned}
x_{n+1}={}&x_n-\frac{\delta t_{\rm int}}{\gamma}
\nabla U_\lambda(x_n)\\
&+\sqrt{\frac{2k_{\rm B}T\delta t_{\rm int}}{\gamma}}\,\xi_n,
\end{aligned}
\label{eq:integrator}
\end{equation}
where $\xi_n\sim\mathcal N(0,I)$, $\delta t_{\rm int}=10^{-4}$~ps and the Cartesian friction is $\gamma=100$~kJ~ps~mol$^{-1}$~nm$^{-2}$. HREX attempts swaps between adjacent states every 1,000 integration steps and saves configurations every ten exchange rounds during production. The fixed-$\lambda$ trajectories used to train the transition model and those used as kinetic references save a frame every 2,000 steps (0.2~ps).

The equilibrium tokeniser is a residual vector-quantized encoder
\cite{vandenOord2017vqvae,zeghidour2022soundstream} that maps all pairwise
atomic distances of a frame to two sequential codebooks of 128 entries. A Cartesian decoder reconstructs the full molecule and is trained with a loss combining aligned-coordinate error, pairwise-distance error, codebook loss and commitment loss, weighted 1, 2, 0.25 and 0.25. The tokenizer is fitted to training frames, and $\phi$ and $\psi$ are computed from decoded coordinates for evaluation. The grid tokens of the transition model are described in Methods.

The equilibrium model factorizes a code pair as $q^{\rm eq}_\theta(a,b\mid\lambda) =q^{\rm eq}_\theta(a\mid\lambda)\,q^{\rm eq}_\theta(b\mid a,\lambda)$. At code position $r\in\{1,2\}$, its logits depend on the coupling through
\begin{equation}
\begin{aligned}
\ell_{r,k}(z_{<r},\lambda)
={}&A_{\theta,r,k}(z_{<r})\\ &+(\lambda-1.075)B_{\theta,r,k}(z_{<r}),
\end{aligned}
\label{eq:affine}
\end{equation}
where 1.075 is the mean of the training couplings, and the same backbone supplies the context-dependent intercept $A$ and slope $B$. The coupling therefore enters through the output heads, without an additional input embedding, and a softmax over each 128-code vocabulary normalizes the logits.

\paragraph{Population calibration.}
The two anchor populations at $\lambda_0=0.9$ and $\lambda_1=1.1$ come from 64 and 32 independent fixed-$\lambda$ chains, respectively, each with 100~ps of burn-in and 3~ns of production at 300~K. For an uncalibrated predictor, let $x_i$ be its predicted basin-1 log odds and $y_i$ the measured MD log odds at anchor $\lambda_i$, where $\operatorname{logit}p=\log[p/(1-p)]$. We set
\begin{equation}
\begin{aligned}
\beta&=\frac{y_1-y_0}{x_1-x_0},
\qquad \alpha=y_0-\beta x_0,\\
\hat p_{\mathrm{cal}}(\lambda)
&=\operatorname{sigmoid}\!\left[\alpha+\beta
\operatorname{logit}\hat p_{\mathrm{raw}}(\lambda)\right],
\end{aligned}
\label{eq:calibration}
\end{equation}
with $\operatorname{sigmoid}(v)=1/(1+e^{-v})$, which is defined whenever the predictor's two anchor log odds differ. The affine map reproduces the two anchor populations exactly but leaves the shape of the prediction between and beyond them to the predictor itself. The map is fitted separately to each of the ten models and to MBAR, using the same two measured populations, and the calibrated model predictions are averaged. The anchor-only baseline is a straight line in log odds through $y_0$ and $y_1$ as a function of $\lambda$, followed by the sigmoid. Intervals come from a 20,000-draw paired bootstrap that resamples the two sets of anchor chains independently and the eight reference chains as complete chains, refitting all maps on each draw.

\paragraph{Frame-isoentropic controller.}
For fixed, finite logits, the mean entropy increases monotonically with the sampling temperature,
\begin{equation}
\frac{d h_s(\tau)}{d\tau}
=\frac{1}{N\tau^3}\sum_{r=1}^{N}
\operatorname{Var}_{p_{s,r}^{\tau}}(\ell_{s,r,\cdot})\geq0,
\label{eq:entropyderivative}
\end{equation}
so the solution of $h_s(\tau_{s+1})=h_\star$ is unique when it exists and is found by bracketing and bisection. During generation of frame $s$, the protein model records the unscaled logits over the 4096 valid codes at each residue along the realised tokens; once the frame is complete, these are held fixed while $\tau_{s+1}$ is solved. All protein results use $M=16$ probes of the first transition to set $h_\star$, except the comparison with the model trained without history, which used $M=1$.

The $\lambda=0$ alanine test applies the same entropy reference to the two-code equilibrium model, which has no learned transition between frames. For an initial first code $a_0$ drawn at $\tau=1$, the target is $h_\star(a_0)=\{H[q_{\theta,1}^{\rm eq}(a\mid\lambda)] +H[q_{\theta,1}^{\rm eq}(b\mid a_0,\lambda)]\}/2$. For each preceding first code $a$, the next-frame temperature $\tau(a)$ solves the same two-term equation with $a$ in place of $a_0$, which defines the controlled kernel
\begin{equation}
K_{a,(a',b')}(\lambda)= q_{\theta,\tau(a)}^{\rm eq}(a'\mid\lambda)\, q_{\theta,\tau(a)}^{\rm eq}(b'\mid a',\lambda).
\label{eq:alaninecontroller}
\end{equation}
We propagate this finite-state kernel exactly for 200 frames, sum the basin-1 occupancy over those frames and average over all initial codes $a_0$, weighted by their model probability at $\tau=1$, which removes the sampling noise of choosing a few starting codes. The resulting chain is an inference construction over frame probabilities, and its 200 steps have no physical duration.

\paragraph{Predictive entropy and likelihood of generated frames.}
Each frame of the five T4L trajectories analysed in Supplementary Results, generated by the history-conditioned model with standard sampling, is passed back through the same model with teacher forcing, conditioned on the sequence, the 320~K temperature token, the lag token and the available earlier frames. At each position, $p_{t,r}(k)=q^{\rm tr}_{\theta,1}(k\mid H_t,c,z_{t+1,<r})$ for the valid structural codes $k$, using the $\tau=1$ logits restricted to the 4096 valid codes. We compute
\begin{equation}
\begin{aligned}
L_t&=-\frac{1}{N}\sum_{r=1}^{N}
\log p_{t,r}(z_{t+1,r}),\\
H_t^{\mathrm{cat}}&=\frac{1}{N}\sum_{r=1}^{N}
\mathcal H(p_{t,r}).
\end{aligned}
\label{eq:nll}
\end{equation}
Because $H^{\rm cat}_t$ is the expected value of $L_t$ under the model's own predictions, the difference $L_t-H^{\rm cat}_t$ measures surprisal beyond that expectation. Associations with structural drift are Spearman correlations computed within each trajectory and summarised by their median across the five trajectories.

\paragraph{Structural comparisons.}
Definitions of the radius of gyration, native contacts and the T4L state assignment are given in Methods. A generated trajectory is counted as departing from and returning to compact conformations when at least three consecutive frames fall at or below the 25th percentile of the reference MD radius of gyration, then at least three at or above its 75th percentile, then at least three at or below the 25th percentile again; the same criterion is applied to the reference MD trajectories. The Trp-cage test uses the K8A variant (sequence \texttt{DAYAQWLADGGPSSGRPPPS}). In the temperature sweep, time-resolved compact occupancy pools the five trajectories within each of ten consecutive 20-frame windows. The extended 320~K analysis scores all 1,000 frames of each trajectory and reports 20-frame compact fractions separately for each (Fig.~\ref{fig:trp}c shows seed 3407, and Supplementary Fig.~\ref{fig:trp1000} all five).

The published T4L RMSD selection contains 99 atoms: the backbone heavy atoms (N, C$\alpha$, C and O) of residues 100 to 120 and the side-chain heavy atoms of Phe114 and Leu133 \cite{vallurupalli2016atomistic}. Because the model generates only N, C$\alpha$ and C, we use the 66 atoms common to both, namely the N, C$\alpha$ and C atoms of residues 100 to 120 and of Leu133. The local RMSD is computed after fitting this subset, and the transition-region placement is the RMSD of the same atoms after global C$\alpha$ alignment of the whole protein \cite{kabsch1976rotation}; this second measurement does not change the state assignment. For the timeline in Fig.~\ref{fig:t4lpath}b, the first 200 frames of seed 3414 under each of the four configurations are assigned against all 95 published structures, and Supplementary Fig.~\ref{fig:t4lwalkers} applies the same procedure to seeds 3412 to 3421.

The torsion-angle random walks perturb the backbone $\phi$ and $\psi$ angles of internal residues, rotating all downstream N, C$\alpha$ and C atoms about the corresponding N--C$\alpha$ or C$\alpha$--C bond. Torsion increments are resampled as $\phi/\psi$ pairs from the 324,000 frame-to-frame changes observed in the ten generated trajectories and applied to a randomly chosen internal residue with a common random sign. At each step, a target consecutive-frame C$\alpha$ RMSD is drawn from the 2,000 pooled displacements of the generated trajectories, pair moves are accumulated until the target is crossed, up to 64 moves, and the final move is scaled by bisection to match the target. Ten walks of 200 frames start from the 3DMV coordinates and are generated from a single random-number stream.

\paragraph{Optimization and reproducibility.}
The alanine tokeniser uses a 64-dimensional latent, two 128-entry codebooks and 1,500 AdamW updates \cite{loshchilov2019adamw} on at most 10,000 training frames. All alanine and protein models load Qwen2.5-14B in 4-bit precision
\cite{dettmers2023qlora} and use rank-64 LoRA with scale 128 and dropout 0.05
on the attention and feed-forward projections, trained with AdamW and weight decay 0.01. Each alanine equilibrium model, including the instance used at $\lambda=0$, is trained for 1,500 updates (six epochs) with batch size 16 and constant learning rates of $5\times10^{-5}$ for the LoRA parameters and $5\times10^{-4}$ for the new token embeddings and both affine heads. The alanine transition model is trained for 540 updates (one epoch) with an effective batch of 8 and learning rates of $5\times10^{-5}$, $5\times10^{-4}$ and $10^{-3}$ for the LoRA parameters, new token embeddings and $\lambda$ projection, with a linear warm-up over 16 updates followed by a constant rate. The alanine models use gradient clipping at 1.0. The protein models use a learning rate of $5\times10^{-5}$ with a linear warm-up, over 8,526 updates for the history-conditioned model and 3,411 for the model trained without history, followed by a constant rate, and an effective batch of 128. The documented history-conditioned training segments ran on 64 AMD Instinct MI355X GPUs at about 2.1~s per optimizer step (about 33~h per epoch), and each alanine model trained on a single MI355X in under 25~minutes.

\clearpage
\balance
\renewcommand{\bibfont}{\footnotesize}
\bibliographystyle{unsrtnat}
\bibliography{md_llm_2_v3}
\end{document}